\documentclass[preprint,groupedaddress]{revtex4-1}
\usepackage{graphicx}
\usepackage{amsmath}
\usepackage{relsize}
\usepackage{hyperref}
\usepackage{multirow}
\usepackage{color}
\usepackage{float}

\begin{document}

\title{London penetration depth and thermal fluctuations in the sulphur hydride 203 K superconductor}





\author{E. F. Talantsev}
\thanks{Correspondence and requests for materials should be addressed to E.F.T. (email:Evgeny.Talantsev@vuw.ac.nz) or to J.L.T. (email: Jeff.Tallon@vuw.ac.nz)}
\affiliation{Robinson Research Institute, Victoria University of Wellington, P.O. Box 33436, Lower Hutt 5046, New Zealand}
\author{W. P. Crump}
\affiliation{Robinson Research Institute, Victoria University of Wellington, P.O. Box 33436, Lower Hutt 5046, New Zealand}
\author{J. G. Storey}
\affiliation{Robinson Research Institute, Victoria University of Wellington, P.O. Box 33436, Lower Hutt 5046, New Zealand}
\affiliation{MacDiarmid Institute for Advanced Materials and Nanotechnology, P.O. Box 33436, Lower Hutt 5046, New Zealand}
\author{J. L. Tallon}
\thanks{Correspondence and requests for materials should be addressed to E.F.T. (email:Evgeny.Talantsev@vuw.ac.nz) or to J.L.T. (email: Jeff.Tallon@vuw.ac.nz)}
\affiliation{Robinson Research Institute, Victoria University of Wellington, P.O. Box 33436, Lower Hutt 5046, New Zealand}
\affiliation{MacDiarmid Institute for Advanced Materials and Nanotechnology, P.O. Box 33436, Lower Hutt 5046, New Zealand}

\date{\today}

\begin{abstract}
Recently, compressed H$_2$S has been shown to become superconducting at 203 K under a pressure of 155 GPa. One might expect fluctuations to dominate at such temperatures. Using the magnetisation critical current, we determine the ground-state London penetration depth, $\lambda_0$=189 nm, and the superconducting energy gap, $\Delta_0$=27.8 meV, and find these parameters are similar to those of cuprate superconductors. We also determine the fluctuation temperature scale, $T_{\textrm{fluc}}=1470$ K, which shows that, unlike the cuprates, $T_c$ of the hydride is not limited by fluctuations. This is due to its three dimensionality and suggests the search for better superconductors should refocus on three-dimensional systems where the inevitable thermal fluctuations are less likely to reduce the observed $T_c$.
\end{abstract}



\maketitle

Motivated by the theoretical prediction by Li {\it et al.} \cite{Li} of high temperature superconductivity in highly-compressed H$_2$S, Drozdov {\it et al.}\cite{Eremets} recently succeeded in demonstrating the onset of superconductivity at the remarkably high temperature of 203 K in H$_2$S at a pressure of 155 GPa.  At such temperatures thermal fluctuations are expected to break Cooper pairs and suppress superconductivity. For cuprates strong fluctuations reduce the observed transition temperature, $T_c$, well below its mean-field value, by up to 30\%. What is their impact on sulphur hydride? Is the mean-field $T_c$ even higher, perhaps 300 K? Do fluctuations impact on the search for room-temperature superconductivity? Here, we determine the fluctuation temperature scale and show that $T_c$ of the hydride is in fact not limited by fluctuations.

For nearly three decades the search for superconductors with higher $T_c$ has concentrated on layered, strongly-correlated systems like the cuprates \cite{Chiao} or pnictides \cite{Stewart}. One of the attractive features here is the potentially high density of electronic states associated with a proximate van Hove singularity \cite{Storey,Bok}. However, because of their quasi-two-dimensional electron dynamics these systems are subject to strong fluctuations at temperatures comparable to $T_c$ and this results in their $T_c$ value being significantly depressed below the mean-field value, which for YBa$_2$Cu$_3$O$_7$ is about 120 K \cite{Tallon1} and for Bi$_2$Sr$_2$CaCu$_2$O$_8$ is inferred to be as high as 150 K \cite{Tallon1,Kondo,Gomes}. Following Emery and Kivelson \cite{Emery} the temperature scale, $T_{fluc}$, for fluctuations in both amplitude and phase \cite{Tallon1} is governed by the superfluid density, $\rho_s \equiv \lambda^{-2}$, according to
\begin{equation}
k_B T_{fluc} =  \frac{A \phi_0^2 a}{4\pi^2 \mu_0 \lambda^2(0)} . \label{fluc}
\end{equation}
\noindent where $\phi_0$ is the flux quantum, $\mu_0$ is the permeability of free space, $k_B$ is Boltzmann's constant and $\lambda(0)$ is the ground-state value of $\lambda(T)$. The parameter, $a$, is a short-range cut-off length which, together with the scale parameter, $A$, are the only elements which capture the effects of dimensionality. For two dimensions (2D) $A$ = 0.9 while for three dimensions (3D) we use their value of $A=2.2$ which is explicitly applicable to the 3D-XY model. Again, following these authors we take $a$ to be the mean spacing between superconducting sheets for 2D systems, while $a^2 = \pi \xi^2(0)$ for 3D, where $\xi$ is the coherence length. (It can be seen that the quantity on the right hand side is also the maximum pinning energy of a quantized flux vortex and therefore $T_{fluc}$ is also related to the in-field critical current density \cite{Tallon2}.) It is the relatively low superfluid density (or large penetration depth) of the cuprates and pnictides that results in $T_{fluc}$ being comparable to $T_c$ and thus ensures the dominant role of fluctuations, so much so as to compromise the possibility of room-temperature superconductivity in this type of system \cite{Tallon3}.

The discovery of superconductivity at 203 K in highly compressed sulphur hydride \cite{Eremets,Troyan} raises the question as to the role of fluctuations in this system and whether the actual mean-field $T_c$ value might already be near 300 K, comparable to that predicted by Ashcroft nearly 50 years ago for compressed hydrogen \cite{Ashcroft,Ashcroft2,Ashcroft3}. If so, all chance of room temperature superconductivity might vanish. We thus sought to determine the fundamental superconducting parameters, $\lambda$ and $\Delta$, of this system in order to extract the value of its fluctuation temperature scale. These parameters are unlikely to be measured by any conventional techniques due to the extreme conditions at which sulphur hydride becomes superconducting and so we use the critical current density to determine their values, as follows.

Recently we showed that the self-field transport critical current density, $J_c(\textrm{sf})$, in thin-film type II superconductors with half-thickness $b \leq \lambda$ is given by \cite{Talantsev}:
\begin{equation}
J_c(T,\textrm{sf}) = \frac{B_{c1}(T)}{\mu_0\lambda(T)} = \frac{\phi_0}{4\pi \mu_0 \lambda^3(T)} \left(\ln\kappa + 0.5\right) . \label{type2}
\end{equation}
Here $B_{c1}$ is the lower critical field and $\kappa = \lambda(0)/\xi(0)$ is the Ginzburg-Landau parameter. We emphasise that this represents a $J_c$ paradigm in which vortices and vortex movement play no role in {\it self-field} $J_c$ \cite{Talantsev,Talantsev2}. Indeed, for very thin samples the surface fields are far smaller than the critical field so that vortices are manifestly not operative, yet Eq~\ref{type2} is still satisfied \cite{Talantsev2}.  Analysis of $J_c$ in terms of Eq.~\ref{type2} thus enables $\lambda(T)$ to be determined and, from its low-$T$ dependence, $\Delta(0)$ to be extracted (Eq. 6, ref.~\cite{Talantsev}). Very recently this analysis has been successfully applied to extract $\lambda$ for ultra-thin films (4-5 nm) of YBa$_2$Cu$_3$O$_7$ \cite{Fete}.

For films with half-thickness $b \geq \lambda$ there is a correction factor to Eq.~\ref{type2} of $\left(\lambda/b\right)\tanh\left(b/\lambda\right)$ which was validated for a number of superconductors \cite{Talantsev}. The method is particularly capable of exposing multiband and/or multiphase behaviour yielding partial band contributions to the superfluid density, the magnitudes of the respective band gaps and/or phase fractions, depending of course on the quality of the $J_c$ data. Here we apply this method to the high-pressure magnetisation data of Drozdov \cite{Eremets} from which these authors extracted $J_c$ using the Bean model \cite{Bean}.

Our method \cite{Talantsev} was established strictly only for {\it transport} $J_c$ and the overall current distribution in a conductor in an applied field when measuring {\it magnetisation} $J_c$ is very different from that supporting a transport current. The zero-field magnetisation $J_c$ extracted from an $M$-$H$ hysteresis loop is not in fact in zero field but there is trapped flux present and, as a result, circulating current density in opposing directions. For these and other reasons the Bean model sometimes overestimates $J_{c,m}(T,B)$ at low field and temperature ($B\approx0$, $T<T_c/2$) \cite{Johansen,Grasso}, which somewhat qualifies our analysis. Despite this we previously used Eq.~\ref{type2} to analyse the magnetization critical current density, $J_{c,m}(T,B\approx0)$, for PrOs$_4$Sb$_{12}$ \cite{Cichorek} and for Rb$_3$C$_{60}$ \cite{Tai}. The derived parameters, $\lambda(0)$ and $\Delta(0)$ are in excellent agreement with reported values \cite{Talantsev}, and the test in the present instance is whether the resulting parameters are reasonable. As we will show that test is met.

With these reservations in mind we proceed with the analysis of the magnetisation data for compressed H$_2$S. In view of the fact that the sample thickness is $\approx 1 \mu$m and after compression will be comparable to $\lambda$ and, moreover, Drozdov {\it et al.} \cite{Eremets} infer a grain radius of $0.1 \mu$m (i.e. $\leq \lambda$) then we apply Eq.~\ref{type2} without any size correction factor to the reported magnetisation $J_c$ data, and thereby calculate $\lambda(T)$ and fit $\lambda(0)$ and $\Delta(0)$.


\begin{figure}
\centerline{\includegraphics*[width=75mm]{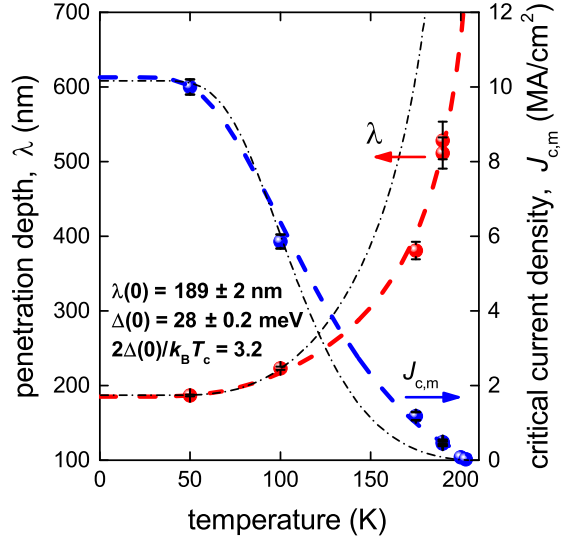}} \caption{\small
$T$-dependence of the magnetisation critical current density (blue data) calculated from $M$-$H$ hysteresis loops \cite{Eremets} as described in the text and penetration depth (red data), calculated using Eq.~\ref{type2}, for H$_2$S at 155 GPa. The black error bars in $J_c$ are the $2\sigma$ errors from reading off the magnetisation data at $H=0$ and these are propagated through to the black error bars shown for $\lambda$. The dashed red curve is the fitted $s$-wave weak-coupling BCS temperature dependence of $\lambda(T)$ and the blue curve is $J_c$ calculated from the red curve using Eq.~\ref{type2}. Black dash-dot curves are their counterparts calculated using the strong-coupling scheme of Gross {\it et al.} \cite{Gross} for the case $2\Delta/(k_B T_c) = 4.5$. }
\label{H2Sfits}
\end{figure}

Fig. 1 shows $J_{c,m}(T,B\approx0)$ calculated from the reported magnetisation $M$-$H$ loops (blue data points, right-hand scale) together with the inferred values of $\lambda(T)$ (red data points) calculated using Eq.~\ref{type2} with, initially, the value of $\kappa$  reported by Drozdov \cite{Eremets}. The error bars represent the 2$\sigma$ random errors obtained by measuring off, and averaging, values of magnetisation at -5 mT, 0 and +5 mT. They do not e.g. reflect any potential systematic errors associated with applying a transport $J_c$ analysis to magnetisation $J_c$ data. Drozdov {\it et al.} estimate $\xi(0)$ from the $T$-dependence of the upper critical field $B_{c2}$. We take $\xi(0)$ = 2.15 nm as the mean value of their reported coherence length, 2.0 - 2.3 nm \cite{Eremets}. Combining this with our deduced $\lambda(0)$ we revise $\kappa$ to recalculate iteratively. The final self-consistent values of $\lambda(T)$ are shown in Fig. 1 (red data points, left-hand scale) and by initially fitting the $s$-wave BCS weak-coupling $T$-dependence of $\lambda$ at low-$T$ we extract $\Delta(0)$ as before \cite{Talantsev} (dashed red curve). It is important to note that the magnitude of $\Delta(0)$ is set by the {\it shape} of the $J_c(T)$ data, not in any way by the absolute value of $\lambda(0)$. Ideally, more low-temperature data is needed to confirm the canonical exponentially-flat $T$-dependence of $\lambda(T)$ at low-$T$ which is characteristic of $s$-wave superconductivity. Nonetheless the overall $s$-wave fit is excellent with a fitting error bar of $\Delta \lambda = \pm 2$ nm. Finally, the blue dashed curve is $J_c$ calculated from the red curve using Eq.~\ref{type2}.

We find $\lambda(0) = 189 \pm 2$ nm, somewhat different from the reported Drozdov value of 125 nm but we note that they inadvertently included a factor of $\sqrt{2}$ in their ratio of $B_{c1}/B_{c2}$. Moreover, we use the factor $(\ln\kappa + 0.5)$ in our definition of $B_{c1}$, not just $\ln\kappa$. Thus, taking the value of $B_{c1} = 30$ mT measured by Drozdov {\it et al.} \cite{Eremets} their reported value becomes $\lambda(0) = 163$ nm, actually much closer now to our inferred value. Moreover, we note that determination of $\lambda$ from direct measurement of $B_{c1}$, as in their case, is notoriously difficult and variable \cite{Talantsev}.

We find from the low-$T$ variation of $\lambda(T)$ that $\Delta(0) = 27.8 \pm 0.2$ meV, so that $2\Delta/(k_B T_c) = 3.2 \pm 0.03$, not much below the weak-coupling BCS value of 3.53 thus indicating more or less conventional BCS behaviour. This contrasts recent reports theoretically suggesting a strong-coupling value $2\Delta/(k_B T_c) \geq 4.5$ \cite{Carbotte,Errea}. As a check we calculated $\lambda(T)$, as described in the Supplementary Material (SM), using the strong-coupling scheme of Gross {\it et al.} \cite{Gross} for the case $2\Delta/(k_B T_c) = 4.5$. The resultant curves for $\lambda(T)$ and for $J_c(T)$ (back-calculated from $\lambda(T)$ using Eq.~\ref{type2}) are shown by the black dash-dot curves. Evidently the reported $J_c$ data appear to be inconsistent with this stronger coupling scenario which reveals the characteristic more rapid suppression of superfluid density nearer $T_c$ due to the onset of pair-breaking arising from strong-coupling. We conclude that even the present sparse $J_c$ data constrains the magnitude of $\Delta(0)$ rather tightly and appears to rule out a strong-coupling scenario. As a further check, the potential impact of the paucity of low-temperature data was tested by suppressing the 50 K data, so that the fit was restricted to data at $T =$ 100 K, 175 K, and 190 K. We obtained nearly the same parameter magnitudes and error bars reported above: $\lambda(0) = 183 \pm 6$ nm and $\Delta(0) = 27.1 \pm 0.3$ meV, thus confirming that our conclusions are not prejudiced by the lack of lower temperature data. As regards the inferred BCS ratio which is lower than the weak-coupling value, such behaviour is not unknown (e.g. Zn \cite{Padamsee}). However we have here used a single-band analysis. In a two-band analysis two gaps will open and in our experience the larger of these gaps will be greater than that obtained in a single-band fit analysis \cite{Talantsev2}. If so then such a larger $\Delta(0)$ value might perhaps be more consistent with a weak-coupling gap ratio of 3.53. To settle this question would require a much more comprehensive data set extending to low temperature.

\begin{table}
\begin{tabular}{|c||c|c|c|c|}
\hline
Material & $a$ (nm) & $\lambda_0$ (nm) & $T_c$ (K) & $T_{fluc}$ (K) \\ \hline
YBa$_2$Cu$_3$O$_y$  & 0.41$^*$ & 128 \cite{Talantsev} & 93 & 140$^{\dag}$ \\ \hline
(Y,Ca)Ba$_2$Cu$_3$O$_y$ & 0.41$^*$ & 133 \cite{Talantsev} & 86 & 130 \\ \hline
Bi$_2$Sr$_2$CaCu$_{2}$O$_8$  & 0.75 & 190 \cite{Talantsev} & 95 & 112 \\ \hline
Ba(Fe,Co)$_2$As$_2$  & 0.65 & 277 \cite{Talantsev} & 40 & 48 \\ \hline
sulphur hydride & 3.8 & 163 \cite{Eremets} & 203 & 1,970 \\
($P$ = 155 GPa) &  & ($B_{c1}$ = 30 mT) & &  \\ \cline{3-5}
  & & 189 (this work) & 203 & 1,470 \\ \hline

\end{tabular}

\caption{\small Parameters used in Eq.~\ref{fluc} to calculate the onset temperature, $T_{fluc}$, for phase fluctuations. $^*$The entry for YBa$_2$Cu$_3$O$_y$ reflects the fact that superconductivity also extends onto the CuO chain layer. $^{\dag}$These are maximal values for $T_{fluc}$: they reduce rapidly in both underdoped and overdoped cuprates \cite{Tallon2}. }
\label{results}
\end{table}

Now, using $\xi(0) = 2.15$ nm in the expression $\Delta(0) = \hbar v_F/(\pi \xi(0))$ we find $v_F = (2.8 \pm0.2) \times 10^5$ m/s, more or less identical to the universal Fermi velocity of the cuprates \cite{Zhou}; and we find the Ginzburg-Landau parameter $\kappa \equiv \lambda(0)/\xi(0) = 88$, thus characterising sulfur hydride as a high-$\kappa$ superconductor. All these parameters are remarkably similar to those of the cuprates and arsenides. However, and in sharp contrast, we find a very large difference in terms of classical thermodynamic phase fluctuations which, as noted in the case of the cuprates, reduce $T_c$ below its mean-field value by up to 30\% \cite{Tallon1,Kondo}. Due to its higher $T_c$ the effects of thermal fluctuations might naively be thought to be even more drastic for sulphur hydride.

It can be seen in Eq.~\ref{fluc} that fluctuations are governed not only by the magnitude of the superfluid density, $\lambda^{-2}$ but also the dimensionality. Using our data for sulphur hydride at 155 GPa we find $T_{fluc} = 1470$ K (see Table 1) in sharp contrast with $T_{fluc}$ = 140 K for YBa$_2$Cu$_3$O$_7$, 112 K \cite{Talantsev} for Bi$_2$Sr$_2$CaCu$_2$O$_8$ \cite{Tallon2} and 48 K for (Ba,K)Fe$_2$As$_2$ \cite{Talantsev}. This huge difference arises solely from the three-dimensionality of sulphur hydride in contrast with the quasi-2D nature of the cuprates and arsenides which, in their case, leads to $T_{fluc}$ values which are comparable to $T_c$ and hence reduces $T_c$ well below its mean-field value \cite{Tallon1}. For sulphur hydride, $T_{fluc} = 1470$ K is comparable to the phonon frequency and is substantially greater than $T_c$ = 203 K. Thus, one expects very weak fluctuations in this system as is evidenced by the rather sharp transitions \cite{Eremets} despite the very high transition temperature which will be very close to the mean-field value.

\begin{table*}
\tiny
\centering
\begin{tabular}{|c|c|c|c|c|c|c|c|c|c|c|}
\hline
$T_c$ & $V_M$ & $\lambda_0$ & $T_{fluc}$ & $\kappa$ & $B_{c1}(0)$ & $B_{c2}(0)$ & $U(0)$ & $\gamma_n$ & $\Delta\gamma_c$ & $N(E_F)$ \\
(K) & (m$^3$/mol) & (nm) & (K) &  & (mT) & (T) & (J/mole) & (mJ/K$^2$/mol) & (mJ/K$^2$/mol) & (states/eV/fu) \\ \hline
203 & $8.8 \times 10^{-6}$ \cite{Eremets2} & $189\pm2$ & $1216\pm97$ & $105\pm8$  & $23.7 \pm 0.4$ & $100 \pm 16$ & $2.25 \pm 0.72$ & $0.299 \pm 0.074$ & $0.427 \pm 0.106$ & $0.063 \pm 0.015$ \\ \hline

\end{tabular}

\caption{Thermodynamic parameters for compressed sulphur hydride at 155 GPa calculated from the critical fields under the assumption of near-weak-coupling BCS behaviour. The zeroes in columns 6-8 refer to ground-state values. }
\label{parameters}
\end{table*}

Additional thermodynamic parameters may be calculated from the available data. To extrapolate $B_{c2}(T)$ to $T = 0$ we do not use the simple idealised quadratic form used by Drozdov {\it et al.} but the Werthamer-Helfand-Hohenberg relation $B_{c2}(0) = -0.693 T_c \times(dB_{c2}/dT)_{T_c}$ which yields a value 39\% higher \cite{Werthamer}. From $B_{c2}(0)$ one may calculate the condensation energy, the normal-state electronic specific heat coefficient, $\gamma_n$, the jump in $\gamma(T)$ at $T_c$, $\Delta\gamma_c$, and the electronic density of states (DOS) at the Fermi level, $N(E_F)$. These calculations are detailed in the SM and the results are summarized here in Table II. (The impact of the higher $B_{c2}(0)$ value on our deduced values of $\lambda(0)$ and $T_{fluc}$ is also discussed in the SM. It is not significant.) The value of $N(E_F)$ in Table II is consistent with the background DOS obtained from electronic band structure calculations \cite{Sano} except that these calculations reveal the presence of a van Hove singularity more or less centred at the Fermi level so that the DOS is significantly energy-dependent. For comparison the specific heat jump, $\Delta\gamma_c$, can in principle be extracted from an analysis of the self-field $J_c$ (see SM) however a much more dense data set, extending to low temperature would be needed.

Finally, one should consider the role of impurity phases in our interpretation. As expected on theoretical grounds \cite{Duan} and very recently confirmed experimentally \cite{Eremets2} highly-compressed H$_2$S dissociates into H$_3$S and elemental sulphur. This raises the inferred $J_c$ which serves only to increase $T_{fluc}$. If, for example, we were to assume the effect is merely to reduce the observed $J_c$ in proportion to the phase fraction then our deduced $\lambda(0)$ should be reduced by a factor of the order of $(2/3)^{(1/3)} = 0.874$ giving $\lambda(0) = 165$ nm, similar to the above-noted value 163 nm obtained from $B_{c1}$ = 30 mT. This yields a yet higher value of $T_{fluc} = 1926$ K and serves only to reinforce our conclusion that it is the three-dimensionality of compressed H$_3$S which ensures the persistence of superconductivity at temperatures above 200 K. What will prove to be most interesting is measurements of $J_c$ to lower temperatures because compressed elemental sulphur itself superconducts up to 14 K and at pressures in excess of 160 GPa converts to another metallic phase with $T_c$ = 17 K \cite{Hemley}. This will be evidenced clearly in $J_c$ studies to low temperature when they are conducted and will enable an independent assessment of phase fractions.

In summary, from magnetisation $J_c$ at low field we determine the fundamental superconducting parameters for sulphur hydride at 155 GPa. We obtain a good fit to the single-band weak-coupling $s$-wave BCS temperature dependence of $\lambda(T)$ with $\lambda(0) = 189 \pm 2$ nm and $\Delta(0) = 27.8 \pm 0.2$ meV. This gives a BCS ratio of $2\Delta/(k_B T_c) = 3.2 \pm 0.03$ suggesting weak coupling, though more low-$T$ data would better confirm this value. Combining this value of $\Delta(0)$ with $\xi(0)$ obtained from $B_{c2}$ measurements we find a Fermi velocity of $v_F = (2.8 \pm0.2) \times 10^5$ m/s. All these values are more or less identical to those of the cuprates yet, in contrast to the cuprates, $T_{fluc}$ = 1470 K greatly exceeds $T_c$ = 203 K so that fluctuations play no role in lowering $T_c$ below its mean-field value. Thus, in addition to the necessary requirement of a high energy scale for the pairing boson, the search for super-high-$T_c$ superconductors should now focus on 3D systems where the inevitable thermal fluctuations are less likely to reduce the observed $T_c$. \\

{\bf Supplementary Material}
The Supplementary Material (SM) describes (i) the calculation of the $T$-dependence of $J_c(\textrm{sf})$ from the $T$-dependence of the superconducting gap, $\Delta(T)$, thus allowing non-linear least squares fitting to extract $\lambda_0$ and $\Delta_0 \equiv \Delta(0)$ in the more general case where coupling strength need not be weak; (ii) the extraction from the data of other thermodynamic parameters including the condensation energy, $U_0$, the normal-state electronic specific heat coefficient, $\gamma_n$, the jump in $\gamma(T)$ at $T_c$, $\Delta\gamma_c$, and the electronic density of states (DOS) at the Fermi level, $N(E_F)$; and (iii) the impact of using the alternative Werthamer-Helfand-Hohenberg formula to extract the value of $B_{c2}$.

The authors thank M. Eremets and N. W. Ashcroft each for reading and commenting helpfully on our manuscript. JLT and JGS each acknowledge separate financial support from the Marsden Fund of New Zealand. EFT acknowledges financial support from Victoria University of Wellington URF Grant No. 8-1620-209864-3580.


\end{document}